\documentclass[preprint,aps]{revtex4}

\usepackage{graphicx}
\usepackage{dcolumn}
\usepackage{bm}
\usepackage{epsf}

\newcommand{\be}{\begin{eqnarray}}

\newcommand{\Eq}[1]{Eq.~(\ref{#1})}

\newcommand{\ur}[1]{(\ref{#1})}

\newcommand{\beq}{\begin{equation}}
\newcommand{\eeq}{\end{equation}}

\newcommand{\la}[1]{\label{#1}}
\newcommand{\bea}{\begin{eqnarray}}
\newcommand{\eea}{\end{eqnarray}}
\newcommand{\beqa}{\begin{eqnarray}}
\newcommand{\eeqa}{\end{eqnarray}}
\newcommand{\ba}{\begin{array}}
\newcommand{\ea}{\end{array}}

\newcommand{\half}{{\textstyle{\frac{1}{2}}}}
\newcommand{\third}{{\textstyle{\frac{1}{3}}}}

\newcommand{\noi}{\noindent}
\newcommand{\n}{\nonumber}

\newcommand{\Th}{$\Theta^+\,$}

\def\appendix{\par
\setcounter{subsection}{0}
\setcounter{equation}{0}

\def\thesection{Appendix}
\def\theequation{\Alph{section}.\arabic{equation}}}

\begin{document}

\title{\bf Exotic pentaquarks as Gamov--Teller resonances}

\author{\bf Dmitri Diakonov$^{1,2}$}

\affiliation{
$^1$Petersburg Nuclear Physics Institute, Gatchina 188300, St. Petersburg, Russia \\
$^2$Institut f\"ur Theoretische Physik II, Ruhr-Universit\"at, Bochum 44780, Germany
}

\date{December 16, 2009}

\begin{abstract}
If the number of colors $N_c$ is taken large, baryons and their excitations can be considered
in a mean-field approach. We argue that the mean field in baryons breaks spontaneously the spherical and $SU(3)$
flavor symmetries, but retains the $SU(2)$ symmetry of simultaneous rotations in space and isospace.
The one-quark and quark-hole excitations in the mean field, together with the $SU(3)$ rotational bands about
them determine the spectrum of baryon resonances, which turns out to be in satisfactory accordance with reality
when one puts $N_c\!=\!3$. A by-product of this scheme is a confirmation of the light pentaquark $\Theta^+$
baryon $uudd\bar s$ as a typical Gamov--Teller resonance long known in nuclear physics. An extension of
the same large-$N_c$ logic to charmed (and bottom) baryons leads to a prediction of a {\it anti-decapenta} ($\overline{15}$)-plet
of charmed pentaquarks, two of which, ${\cal B}_c^{++}=cuud\bar s$ and ${\cal B}_c^+=cudd\bar s$,
may be light and stable with respect to strong decays, and should be looked for~\cite{footnote-1}.\\



\end{abstract}

\maketitle

\section{Relativistic mean field}

It has been argued 30 years ago by Witten~\cite{Witten-Nc} that if the number of colors $N_c$ is large,
the $N_c$ quarks of a baryon can be viewed as moving in a mean field. It is helpful
to understand how baryons look like in the large-$\!N_c$ limit, before $1/N_c$ corrections are considered.

At the microscopic level quarks experience only color interactions, however large $N_c$ do not suppress
gluon fluctuations: the mean field can be only `colorless'. An example how originally color interactions are
Fierz-transformed into interactions of quarks with mesonic fields are provided by the instanton liquid model~\cite{D-02}.

We shall thus assume that quarks in the large-$N_c$ baryon obey the Dirac equation in a background mesonic
field since there are no reasons to expect quarks to be non-relativistic, especially in excited baryons.
In a most general case the background field couples to quarks through all five Fermi variants.
If the background field is stationary in time, it leads to the eigenvalue equation for the $u,d,s$ quarks
in the background field:
\bea\n
H\psi &=& E\psi,\\
H &=& \gamma^0\!\left(\!-i\partial_i\gamma^i+S({\bf x})+P({\bf x})i\gamma^5
+V_\mu({\bf x})\gamma^\mu+A_\mu({\bf x})\gamma^\mu\gamma^5+T_{\mu\nu}({\bf x})\frac{i}{2}[\gamma^\mu\gamma^\nu]\!\right),
\la{DiracH}\eea
where $S,P,V,A,T$ are the mean fields that are matrices in flavor. In fact, the one-particle Dirac Hamiltonian
\ur{DiracH} is generally nonlocal, however that does not destroy symmetries in which we are primarily interested.
We include dynamically-generated quarks masses into the scalar term $S$.

The key issue is the symmetry of the mean field. From the large-$N_c$ point of view, the current strange
quark mass is very small, $m_s={\cal O}(1/N_c^2)$~\cite{D-09}, therefore a good starting point is exact $SU(3)$
flavor symmetry. A natural assumption, then, would be that the mean field is flavor-symmetric, and
spherically symmetric. This assumption, however, leads to too many ``missing resonances'' in the spectrum.
In addition, we know that baryons are strongly coupled to pseudoscalar mesons ($g_{\pi NN}\approx 13$).
It means that there is a large pseudoscalar field inside baryons; at large $N_c$ it is a classical mean field.
There is no way of writing down the pseudoscalar field that would be compatible with the
$SU(3)_{\rm flav}\times SO(3)_{\rm space}$ symmetry. The minimal extension of spherical symmetry is to write
the ``hedgehog'' {\it Ansatz} ``marrying'' the isotopic and space axes:
\beq
\pi^a({\bf x})=\left\{\begin{array}{ccc}n^a\,F(r),& n^a=\frac{x^a}{r},& a=1,2,3,\\
0,&&a=4,5,6,7,8.\end{array}\right.
\la{hedgehog}\eeq
This {\it Ansatz} breaks the $SU(3)_{\rm flav}$ symmetry. Moreover, it breaks the symmetry under
independent space $SO(3)_{\rm space}$ and isospin $SU(2)_{\rm iso}$ rotations, and only a simultaneous
rotation in both spaces remains a symmetry, since a rotation in the isospin space labeled by $a$,
can be compensated by the rotation of the space axes. Therefore, the {\it Ansatz} \ur{hedgehog} breaks
spontaneously the original $SU(3)_{\rm flav}\times SO(3)_{\rm space}$ symmetry down to the
$SU(2)_{{\rm iso}\!+\!{\rm space}}$ symmetry. It is analogous to the spontaneous breaking
of spherical symmetry by the ellipsoid form of many nuclei.

\section{Quarks in the `hedgehog' mean field}

We shall call the $SU(2)_{{\rm iso}\!+\!{\rm space}}$ symmetry of the mean field the ``hedgehog symmetry''.
What mesonic fields $S,P,V,A,T$ in \Eq{DiracH} are compatible with this symmetry? Since $SU(3)$
symmetry is broken, all fields can be divided into three categories:

\noi\underline{I. Isovector fields acting on $u,d$ quarks}
\vskip -0.5true cm

\bea\la{ud-isovector}
{\rm pseudoscalar:}\; P^a({\bf x})\!\!&=&\!\!n^a\,P_0(r),\\
\n
{\rm vector:}\; V^a_i({\bf x})\!\!&=&\!\!\epsilon_{aik}\,n_k\,P_1(r),\\
\n
{\rm axial:}\; A^a_i({\bf x})\!\!&=&\!\!\delta_{ai}\,P_2(r)+n_an_i\,P_3(r),\\
\n
{\rm tensor:}\; T^a_{ij}({\bf x})\!\!&=&\!\!\epsilon_{aij}\,P_4(r)+\epsilon_{bij}\,n_an_b\,P_5(r).
\eea
\noi\underline{II. Isoscalar fields acting on $u,d$ quarks}
\vskip -0.5true cm

\bea\la{ud-isoscalar}
{\rm scalar:}\quad S({\bf x})\!\!&=&\!\!Q_0(r),\\
\n
{\rm vector:}\quad V_0({\bf x})\!\!&=&\!\!Q_1(r),\\
\n
{\rm tensor:}\quad T_{0i}({\bf x})\!\!&=&\!\!n_i\,Q_2(r).
\eea

\noi\underline{III. Isoscalar fields acting on $s$ quarks}
\vskip -0.5true cm

\bea\la{s-isoscalar}
{\rm scalar:}\quad S({\bf x})\!\!&=&\!\!R_0(r),\\
\n
{\rm vector:}\quad V_0({\bf x})\!\!&=&\!\!R_1(r),\\
\n
{\rm tensor:}\quad T_{0i}({\bf x})\!\!&=&\!\!n_i\,R_2(r).
\eea
All the rest fields and components are zero as they do not satisfy the $SU(2)$ symmetry and/or the needed discrete
$C,P,T$ symmetries. The 12 `profile' functions $P_{0,1,2,3,4,5},\,Q_{0,1,2}$ and $R_{0,1,2}$ should be eventually
found self-consistently from the minimization of the mass of the ground-state baryon. However, even if we do
not know those profiles, there are important consequences of this {\it Ansatz} for the baryon spectrum.


Given the {\it Ansatz},
the Hamiltonian \ur{DiracH} actually splits into two: one for $s$ quarks and the other
for $u,d$ quarks. The former commutes with the angular momentum of $s$ quarks,
${\bf J}={\bf L}+{\bf S}$, and with the inversion of spatial axes, hence all energy levels
are characterized by half-integer $J^P$ and are $(2J+1)$-fold degenerate.
The latter commutes only with the `grand spin' ${\bf K}={\bf T}+{\bf J}$ and with inversion,
hence the $u,d$ quark levels have definite integer $K^P$ and are $(2K+1)$-fold degenerate.
The energy levels for $u,d$ quarks on the one hand and for $s$ quarks on the other are
completely different, even in the chiral limit $m_s\to 0$.

All energy levels, both positive and negative, are probably discrete owing to confinement.
Indeed, a continuous spectrum would correspond to a situation when quarks are free at large
distances from the center, which contradicts confinement. [One can model confinement
by forcing the effective quark masses to grow at infinity, {\it e.g.}
$Q_0({\bf x})\sim R_0({\bf x})\sim \sigma r$.]

According to the Dirac theory, all {\em negative}-energy levels, both for $s$ and $u,d$ quarks,
have to be fully occupied, corresponding to the vacuum. It means that there must be exactly
$N_c$ quarks antisymmetric in color occupying all (degenerate) levels with $J_3$ from $-J$ to $J$,
or $K_3$ from $-K$ to $K$; they form closed shells that do not carry quantum numbers.
Filling in the lowest level with $E>0$ by $N_c$ quarks makes a baryon~\cite{DPP-88,D-09}, see Fig.~1.

\begin{figure}[htb]
\begin{minipage}[t]{.45\textwidth}
\includegraphics[width=\textwidth]{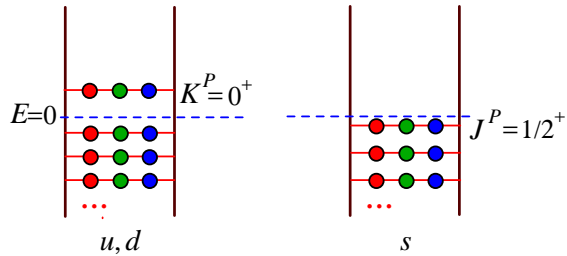}
\end{minipage}
\caption{Filling $u,d,s$ shells for the ground-state baryons: $({\bf 8},{1/2}^+),\;({\bf 10},{3/2}^+)$.}
\label{fig:1}
\end{figure}

The mass of a baryon is the aggregate energy of all filled states, and being a functional
of the mesonic field it is proportional to $N_c$ since all quark levels are degenerate in color.
Therefore quantum fluctuations of mesonic field in baryons are suppressed as $1/N_c$ so that
the mean field is indeed justified.

Quantum numbers of the lightest baryons are determined from the quantization of the rotations
of the mean field, leading to specific $SU(3)$ multiplets that reduce at $N_c\!=\!3$ to the octet
with spin $\half$ and the decuplet with spin $\frac{3}{2}\;$, see {\it e.g.}~\cite{DP-08}.
Witten's quantization condition $Y'\!=\!\frac{N_c}{3}$~\cite{Witten-Sk} follows trivially from
the fact that there are $N_c$ $\;u,d$ valence quarks each with the hypercharge $\frac{1}{3}$~\cite{Blotz}.
Therefore, the ground state shown in Fig.~1 entails in fact 56 rotational states.
The splitting between the centers of the multiplets $({\bf 8},\half^+)$ and $({\bf 10}, \frac{3}{2}^+)$
is ${\cal O}(1/N_c)$, and the splittings inside multiplets can be determined as a perturbation
in $m_s$~\cite{Blotz}.

\section{Excited states in the mean field}

The lowest baryon resonance beyond the rotational excitations of the ground state
is the singlet $\Lambda(1405,\half^-)$. Apparently, it can be obtained only as an excitation
of the $s$ quark, and its quantum numbers must be $J^P=\half^-$~\cite{D-09}, see transition
{\it 1} in Fig.~2.

The existence of an $\half^-$ level for $s$ quarks automatically implies that there is a
particle-hole excitation of this level by an $s$ quark from the $\half^+$ level.
We identify this transition {\it 2} with $N(1535,\half^-)$~\cite{D-09}.
It is predominantly a pentaquark state $u(d)uds\bar s$ (at $N_c\!=\!3$).
This explains its large branching ratio in the $\eta N$ decay~\cite{Zou}, a long-time mystery.
We also see that, since the highest filled level for $s$ quarks is lower than the highest filled
level for $u,d$ quarks, $N(1535,\half^-)$ must be {\em heavier} than $\Lambda(1405,\half^-)$:
the opposite prediction of the non-relativistic quark model has been always of some concern.
Subtracting $1535-1405=130$, we find that the $\half^+$ $s$-quark level is approximately
130 MeV lower in energy than the valence $0^+$ level for $u,d$ quarks.

\begin{figure}[htb]
\begin{minipage}[t]{.45\textwidth}
\includegraphics[width=\textwidth]{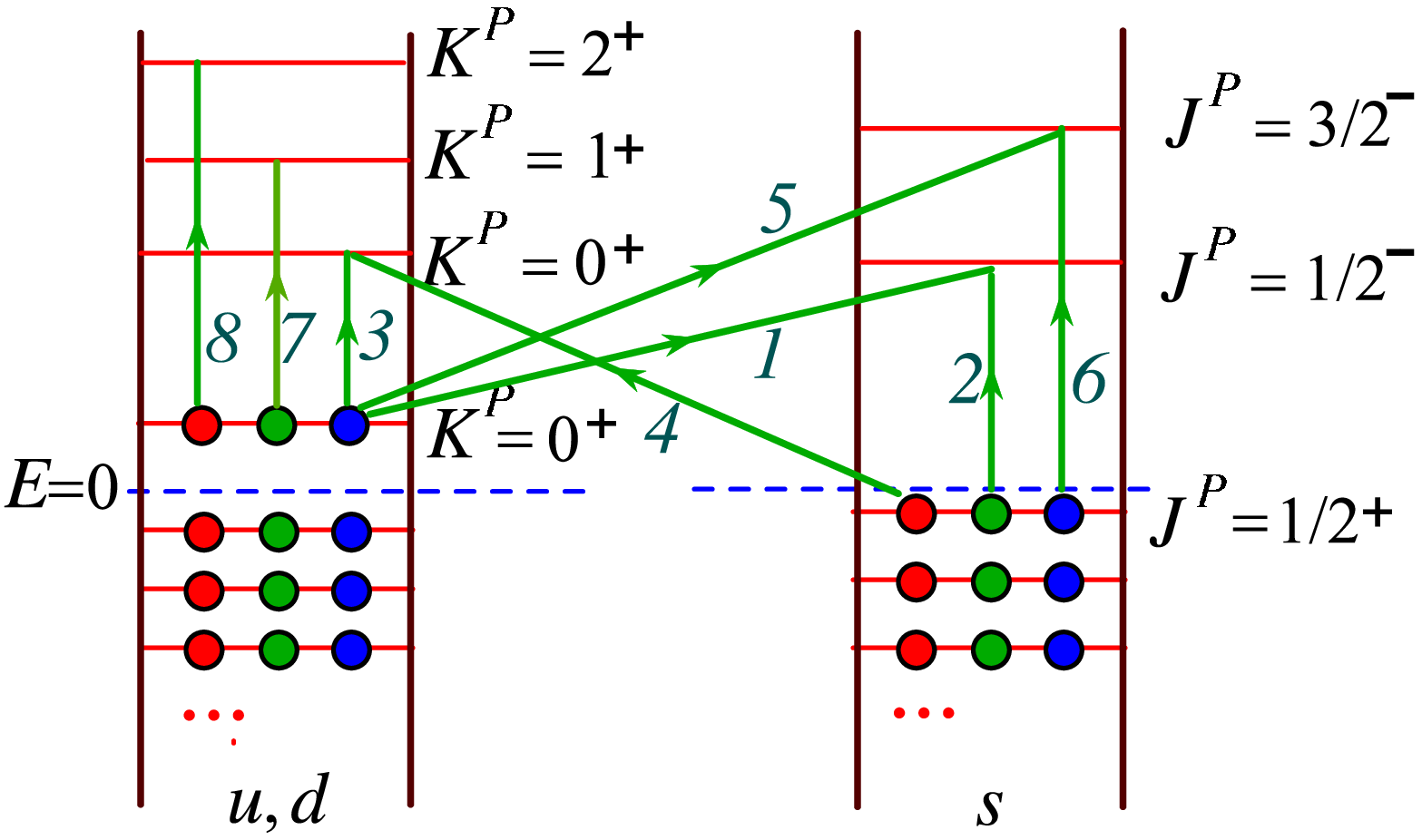}
\end{minipage}
\caption{All baryon resonances below 2 GeV follow from this scheme of one-quark levels.
The transitions shown by arrows correspond to:
{\it 1}: $\Lambda(1405,{1/2}^-)$,
{\it 2}: $N(1535,{1/2}^-)$,
{\it 3}: $\!\!N(1440,{1/2}^+\!)\!$,
{\it 4}: $\!\!\Theta^+(1530,{1/2}^+\!)\!$,
{\it 5}: $\!\!\Lambda(1520,{3/2}^-\!)\!$,
{\it 6}: $\!\!N(1650,{1/2}^-?\!)\!$,
{\it 7}: $\!\!N(1710,{3/2}^+\!)\!$,
{\it 8}: $\!\!N(1680,{5/2}^+\!)\!$. Other resonances belong to $SU(3)$ multiplets
obtained as rotational excitations of these one-particle and particle-hole excitations.}
\label{fig:2}
\end{figure}

The low-lying Roper resonance $N(1440,\half^+)$ requires
an excited one-particle $u,d$ state with $K^P=0^+$~\cite{D-09}, see transition {\it 3}.
Just as the ground state nucleon, it is part of the excited $({\bf 8'},\half^+)$
and $({\bf 10'}, \frac{3}{2}^+)$ split as $1/N_c$.
Such identification of the Roper resonance
solves another problem of the non-relativistic model where $N(1440,\half^+)$ must be heavier
than $N(1535,\half^-)$. In our approach they are unrelated.

Given that there is an excited $0^+$ level for $u,d$ quarks, one can put there an $s$ quark as well,
taking it from the $s$-quark $\half^+$ shell, see transition {\it 4}.
It is a particle-hole excitation with the valence $u,d$ level left untouched, its quantum numbers being
$S=+1,\;T=0,\;J^P=\half^+$. At $N_c\!=\!3$ it is a pentaquark state $uudd\bar s$, precisely
the exotic $\Theta^+$ baryon predicted in Ref.~\cite{DPP-97} from other considerations.
The quantization of its rotations produces
the antidecuplet $({\bf \overline{10}},\half^+)$.
In our original prediction the ${\cal O}(1)$ gap between $\Theta^+$ and the nucleon was due to the
rotational energy only, whereas here the main ${\cal O}(1)$ part of that gap is due to the one-particle
levels, while the rotational energy is ${\cal O}(1/N_c)$. Methodologically, it is more satisfactory.

In nuclear physics, excitations generated by the axial current $j_{\mu\,5}^\pm$,
when a neutron from the last occupied shell is sent to an unoccupied proton level or {\it v.v.}
are known as Gamov--Teller transitions~\cite{BM}. Thus our interpretation of the \Th is that it is a
Gamov--Teller-type resonance long known in nuclear physics.

An unambiguous feature of our picture is that {\bf the exotic pentaquark is a consequence of the three
well-known resonances and must be light.} Indeed, the $\Theta^+$ mass can be estimated from the sum rule~\cite{D-09}:
$m_{\Theta}\approx 1440+1535-1405\approx 1570\,{\rm MeV}$, however there are ${\cal O}(m_s)$
corrections to this equation. 

To account for higher baryon resonances one has to assume that there
are higher one-particle excitations, both in the $u,d$- and $s$-quark sectors, shown in Fig.~2.
It is easy to obtain that order of levels under mild assumptions about the profile
functions \ur{ud-isovector}--\ur{s-isoscalar}.

\section{Baryon resonances from rotational bands}

The original $SU(3)_{\rm flav}\times SO(3)_{\rm space}$ symmetry is restored when
flavor and space rotations are accounted for. Each transition in Fig.~2 generally entails ``rotational bands'' of $SU(3)$
multiplets with definite spin and parity. The short recipe of getting them is: Find the hypercharge
$Y'$ from the number of $u,d,s$ quarks involved; only those multiplets are allowed that contain this $Y'$.
Take an allowed multiplet and read off the isospin(s) $T'$ of particles at this value of $Y'$.
The allowed spin of the multiplet obeys the angular momentum addition law:
${\bf J}={\bf T'}+{\bf J_1}+{\bf J_2}+{\bf K_1}+{\bf K_2}$ where $J_{1,2}$ and $K_{1,2}$ are the initial
and final momenta of the $s$ and $u,d$ shells involved in the transition, respectively. The mass of the
center of a multiplet does not depend on ${\bf J}$ but only on ${\bf T'}$ according to the relation~\cite{DP-04}
\beq\n
{\cal M}={\cal M}_0+\frac{C_2(p,q)-T'(T'+1)-\frac{3}{4}{Y'}^2}{2I_2}+\frac{T'(T'+1)}{2I_1}
\eeq
where $C_2(p,q)=\third(p^2+q^2+pq)+p+q$ is the quadratic Casimir eigenvalue of the multiplet,
$I_{1,2}={\cal O}(N_c)$ are moments of inertia. After the rotational band for a given transition is constructed,
one has to check if the rotational energy of a particular multiplet is ${\cal O}(1/N_c)$ and not ${\cal O}(1)$,
and if it is compatible with Fermi statistics at $N_c\!=\!3$: some {\it a priori} possible multiplets drop out.
One gets a satisfactory description of all baryon resonances up to about 2 GeV, to be published separately.

\section{Charmed and bottom baryons}

If one of the $u,d$ quarks in a light baryon is replaced by a heavy $b$ or $c$ quark,
there are still $N_c\!-\!1$ $u,d$ quarks left. At large $N_c$, they form {\em the same} mean field as
in light baryons, with the same sequence of Dirac levels (up to $1/N_c$ corrections). The heavy quark
contributes to the mean $SU(3)$-symmetric field but it is a $1/N_c$ correction, too.

The filling of Dirac levels for the ground-state $c$ (or $b$) baryon is shown in Fig.~3:
there is a hole in the $0^+$ shell for $u,d$ quarks. Quantizing rotations of this state
leads to the following $SU(3)$ multiplets: $({\bf \bar 3},{1/2}^+)$, $({\bf 6},{1/2}^+)$ and $({\bf 6},{3/2}^+)$.
The last two are degenerate whereas the first is split from the rest by ${\cal O}(1/N_c)$. The splitting
inside multiplets is ${\cal O}(m_sN_c)$.

\begin{figure}[htb]
\begin{minipage}[t]{.45\textwidth}
\includegraphics[width=\textwidth]{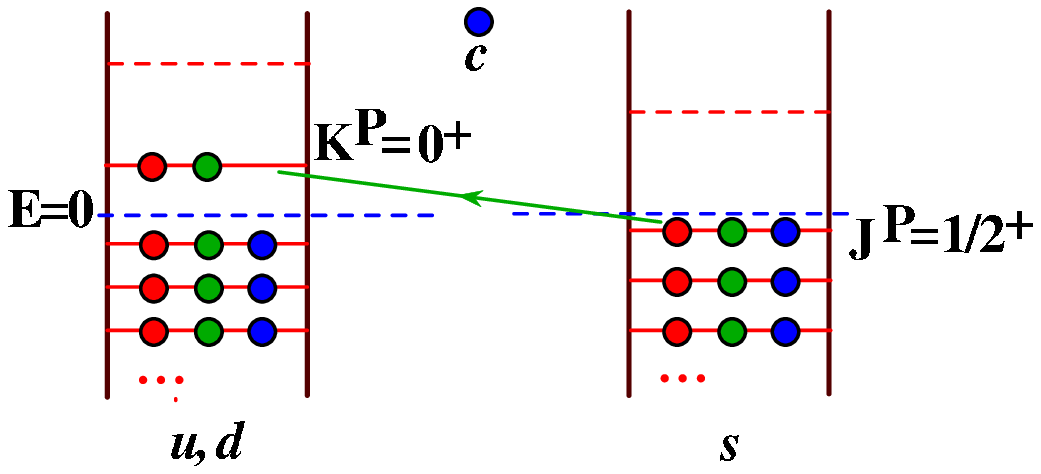}
\end{minipage}
\caption{Filling $u,d,s$ shells for the ground-state charmed baryons,
$({\bf \bar 3},{1/2}^+)$, $({\bf 6},{1/2}^+)$ and $({\bf 6},{3/2}^+)$.
The arrow shows the Gamov--Teller excitation leading to charmed pentaquarks forming $({\bf\overline{15}},{1/2}^+)$.}
\label{fig:3}
\end{figure}

There are good candidates for those ground-state multiplets: $\Lambda_c(2287)$ and $\Xi_c(2468)$ for
$({\bf \bar 3},{1/2}^+)$; $\Sigma_c(2455)$, $\Xi_c(2576)$ and $\Omega_c(2698)$ for $({\bf 6},{1/2}^+)$;
finally $\Sigma_c(2520)$, $\Xi_c(2645)$ and $\Omega_c(2770)$ presumably form $({\bf 6},{3/2}^+)$.
There are ${\bf\bar 3}$'s and ${\bf 6}$'s with parity minus arising from exciting the ${1/2}^-$ $s$-quark level.
The lightest are the degenerate singlets, presumably $\Lambda_c(2595,{1/2}^-)$ and $\Lambda_c(2625,{3/2}^-?)$.

Our new observation is that there is a Gamov--Teller-type transition
when axial current annihilates a strange quark in the $\half^+$ shell, and creates an
$u$ or $d$ quark in the $0^+$ shell, like in the case of the $\Theta^+$. In heavy baryons it is even more simple
as there is a hole in the $u,d$ $0^+$ valence shell from the start. Filling in this
hole means making charmed pentaquarks which we name ``beta baryons'', ${\cal B}_c^+=cudd\bar s$
and ${\cal B}_c^{++}=cuud\bar s$.
Quantizing rotations tells us that these pentaquarks are members of
the {\it anti-decapenta-plet} $({\bf \overline{15}},{1/2}^+)$, Fig.~4. In fact, there must be two additional (nearly
degenerate) multiplets, one with spin ${1/2}^+$ and the other with spin ${3/2}^+$.

Charmed pentaquarks have been considered by Wu and Ma in another approach~\cite{WuMa}; however, they get far larger masses
and in addition pentaquarks with $\bar c$ quarks appear almost degenerate with
those made of $c$ quarks. In our picture the lightest $\bar c$ pentaquarks $\Theta_c$ probably
arise from putting the fourth ($s$) quark at the $\half^-$ level; they form a quadruplet, have parity minus, and are much heavier.

\begin{figure}[htb]
\begin{minipage}[t]{.45\textwidth}
\includegraphics[width=\textwidth]{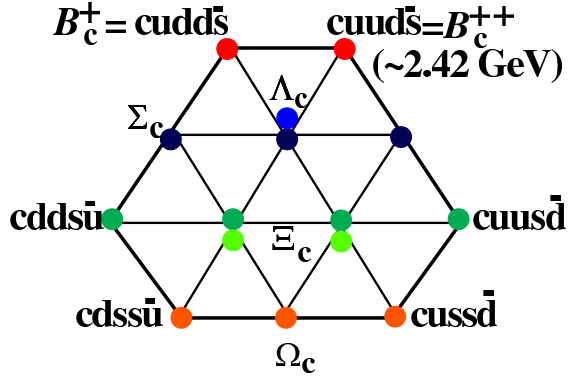}
\end{minipage}
\caption{Decapenta-plet of charmed pentaquarks.}
\label{fig:3}
\end{figure}

Since we know the separation between the ${1/2}^+$ level for $s$ quarks and the $0^+$ level for
$u,d$ quarks from fitting the light baryon resonances, and assuming that it does not change for heavy baryons
(as it would be at $N_c\to\infty$), we estimate the mass of the ${\cal B}_c^{++,+}$ pentaquarks
at about 2420 MeV! The corresponding bottom pentaquarks are about $m(\Lambda_b)+130\,{\rm MeV}=5750\,{\rm MeV}$.
Such light charmed and bottom pentaquarks have no strong decays. Their weak decays, for example
${\cal B}_c^+ \to p\phi\to pK^+K^-$, have clear signatures especially in a vertex detector, and should be
looked for at LHC, Fermilab and B-factories. A cautionary remark, though, is that the production rate is expected to be
quite low.

A detailed elaboration of the ideas presented here will be published elsewhere.\\

I am grateful to Victor Petrov, Maxim Polyakov and Alexei Vladimirov for their help.
I thank Ben Mottelson and Semen Eidelman for useful discussions and Harry Lipkin for a correspondence.
This work has been supported in part by Russian Government grants RFBR-06-02-16786 and RSGSS-3628.2008.2,
and by Mercator Fellowship (DFG, Germany).


\begin{thebibliography}{99}



\bibitem{footnote-1}
An extended version of the invited talk at {\it Quark Nuclear Physics - 2009}, Beijing, Sep. 21-26, 2009,
to be published in Chinese Physics C.

\bibitem{Witten-Nc}
E.~Witten, Nucl. Phys. {\bf B160} (1979) 57.

\bibitem{D-02}
D.~Diakonov, Prog. Part. Nucl. Phys. {\bf 51} (2003) 173, arXiv:hep-ph/0212026.

\bibitem{D-09}
D.~Diakonov, JETP Letters {\bf 90} (2009) 407 [Pis'ma v ZHETF, {\bf 90} (2009) 451], arXiv:0812.3418 [hep-ph];
{\it Nucl. Phys.} {\bf A827} (2009) 264C, arXiv:0901.1373 [hep-ph].

\bibitem{DPP-88}
D.~Diakonov, V.~Petrov and P.~Pobylitsa, Nucl. Phys. {\bf B306} (1988) 809.

\bibitem{DP-08}
D.~Diakonov and V.~Petrov, arXiv:0812.1212 [hep-ph], to be published in {\it The Multifaceted Skyrmion}, G.~Brown and M.~Rho, eds.,
World Scientific.

\bibitem{Witten-Sk}
E.~Witten, Nucl. Phys. {\bf B223} (1983) 433.

\bibitem{Blotz}
A.~Blotz, D.~Diakonov, K.~Goeke, N.W.~Park, V.~Petrov and P.~Pobylitsa, Nucl. Phys. {\bf A355} (1993) 765.

\bibitem{Zou}
B.-S.~Zou, Eur. Phys. J.  {\bf A35} (2008) 325, arXiv:0711.4860 [nucl-th].

\bibitem{DPP-97}
D.~Diakonov, V.~Petrov and M.~Polyakov, Zeit. Phys. {\bf A359} (1997) 305, arXiv:hep-ph/9703373.

\bibitem{DP-04}
D. Diakonov and V. Petrov, Phys. Rev {\bf D69} (2004), 056002, arXiv:hep-ph/0309203.

\bibitem{BM}
A.~Bohr and B.~Mottelson. Nuclear structure. New York: W.~A.~Benjamin (1998) vol. 1.

\bibitem{WuMa}
B.~Wu and B.-Q. Ma, Phys. Rev. {\bf D70} (2004) 034025, arXiv: hep-ph/0402244.


\end{thebibliography}
\end{document}